\pdfoutput=1
 
\documentclass[PAPER,atlasdraft=true,cernpreprint=true,txfonts=true,texlive=2016,USenglish,texmf]{atlasdoc}
 
\usepackage{atlaspackage}
 
\usepackage{atlasbiblatex}
 
\usepackage[misc=true,hion=true]{atlasphysics}
 
\graphicspath{{logos/}{figures/}}
 
\usepackage{lineno}
\usepackage{bm}

\newcommand{\Eref}{\eta_{\mathrm{ref}}}
\newcommand{\lr}[1]{\left\langle #1\right\rangle}
\newcommand{\npart}{\mbox{$N_{\mathrm{part}}$}}

\newcommand*{\HIJING}{\textsc{Hijing}\xspace}

\AtlasTitle{Longitudinal flow decorrelations in Xe+Xe collisions at
$\sqrt{s_{\mathrm{NN}}}=5.44$~$\TeV$ with the ATLAS detector}
 
\date{\today}
 
\AtlasJournal{Phys.\ Rev.\ Letts.}
 
\PreprintIdNumber{CERN-EP-2019-275}
\AtlasJournalRef{Phys. Rev. Lett. 126 (2021) 12230}
\AtlasDOI{10.1103/PhysRevLett.126.122301}

\AtlasAbstract{
The first measurement of longitudinal decorrelations of harmonic flow amplitudes
$v_n$ for $n=2$, 3 and 4 in Xe+Xe collisions at $\sqrt{s_{\mathrm{NN}}}=5.44$~$\TeV$ is obtained using
3~${\mathalpha{\mu}}\textrm{b}^{-1}$ of data with the ATLAS detector at the LHC. The decorrelation
signal for $v_3$ and $v_4$ is found to be nearly independent of collision centrality and transverse
momentum ($p_{\mathrm{T}}$) requirements on final-state particles, but for $v_2$ a strong centrality
and $p_{\mathrm{T}}$ dependence is seen. When compared with the results from Pb+Pb collisions at
$\sqrt{s_{\mathrm{NN}}}=5.02$~$\TeV$, the longitudinal decorrelation signal in mid-central Xe+Xe
collisions is found to be larger for $v_2$, but smaller for $v_3$. Current hydrodynamic models
reproduce the ratios of the $v_n$ measured in Xe+Xe collisions to those in Pb+Pb collisions but fail
to describe the magnitudes and trends of the ratios of longitudinal flow decorrelations between
Xe+Xe and Pb+Pb. The results on the system-size dependence provide new insights and an important lever-arm to separate effects of the longitudinal structure of the initial state from other early-time and late-time effects in heavy-ion collisions.}
\hypersetup{pdftitle={ATLAS document},pdfauthor={The ATLAS Collaboration}}
 
\begin{document}
 
\maketitle
 
High-energy heavy-ion collisions create a new state of matter known as a quark--gluon plasma (QGP), whose space-time dynamics is well described by relativistic viscous hydrodynamic models~\cite{Gale:2013da,Heinz:2013th,Jia:2014jca}. During its expansion, the large pressure gradients of the QGP convert the spatial anisotropies in the initial-state geometry into momentum anisotropies of the final-state particles. Such momentum anisotropies are often characterized by a Fourier expansion of particle density in the azimuthal angle $\phi$, $dN/d\phi\propto 1+2\sum_{n=1}^{\infty} v_n\cos\,n(\phi-\Phi_n)$, where $v_n$ and $\Phi_n$ are the magnitude and phase of the $n^{\mathrm{th}}$-order flow vector $V_n=v_n e^{-{\textrm{i}}n\Phi_n}$. The $V_n$ reflects the hydrodynamic response of the QGP to the shape of the overlap region in the transverse plane, described by eccentricity vector ${\mathcal{E}}_n=\varepsilon_n e^{-{\textrm i}n\Psi_{n}}$~\cite{Niemi:2012aj}. Extensive studies of $V_n$ and their event-by-event fluctuations in a broad range of beam energy and collision systems~\cite{Adare:2011tg,ALICE:2011ab,Aad:2012bu,Chatrchyan:2013kba,Aad:2013xma,Aad:2014fla,Aad:2015lwa,ALICE:2016kpq,Sirunyan:2017fts,Acharya:2018lmh,Adam:2019woz} have provided strong constraints on the ${\mathcal{E}}_n$ and the properties of the QGP~\cite{Luzum:2012wu,Teaney:2010vd,Gale:2012rq,Niemi:2012aj,Qiu:2012uy,Teaney:2013dta}.
 
Most previous efforts assume that the shape of the initial overlap and dynamic evolution of the QGP are boost-invariant. Recently, LHC experiments made the first observation of ``flow decorrelations'' in Pb+Pb collisions~\cite{Khachatryan:2015oea,Aaboud:2017tql}, which show that, even in a single event, $v_n$ and $\Phi_n$ can fluctuate along the longitudinal direction. This can be attributed to the fact that the distribution of particle production sources, and the associated eccentricity vectors, fluctuates along pseudorapidity ($\eta$). For example, the number of forward-going and backward-going nucleon participants, and the corresponding eccentricity vectors ${\mathcal{E}}^{\mathrm{F}}_n$ and ${\mathcal{E}}^{\mathrm{B}}_n$ are not the same in a given event. While the harmonic flow $V_n$ are driven by the average of the two eccentricity vectors $V_n\propto {\mathcal{E}}_{n} \approx \frac{{\mathcal{E}}_n^{\mathrm F}+{\mathcal{E}}_n^{\mathrm B}}{2}$, the flow decorrelation is related to the difference between them, ${\mathcal{E}}_{n-} = \frac{{\mathcal{E}}_n^{\mathrm F}-{\mathcal{E}}_n^{\mathrm B}}{2}$~\cite{Jia:2017kdq}. Indeed, hydrodynamic model and transport model calculations~\cite{Bozek:2010vz,Jia:2014ysa,Bozek:2015bna,Pang:2015zrq,Shen:2017bsr,Bozek:2017qir} show that the flow decorrelations are driven mostly by longitudinal fluctuation of ${\mathcal{E}}_n$ in the initial-state geometry. They are also influenced by other early-time effects, such as initial-state momentum anisotropy~\cite{Pang:2018zzo} and hydrodynamic fluctuations~\cite{Sakai:2020pjw}, but are insensitive to late-time dynamics including shear viscosity~\cite{Pang:2015zrq}. These different early-time contributions compete with each other, and current measurements~\cite{Khachatryan:2015oea,Aaboud:2017tql} from a single system (Pb+Pb) in a limited energy range ($\sqn=2.76$--5.02 $\TeV$) do not disentangle these effects. To improve our understanding of the longitudinal structure of the QGP, it is crucial to extend the measurements to a broad range in the beam energy and size of the collision systems~\cite{Pang:2015zrq,Behera:2020mol}.
 
This Letter investigates the system-size dependence of longitudinal decorrelations of $v_2$, $v_3$, and $v_4$ by performing measurements in $^{129}$Xe+$^{129}$Xe collisions and comparing them with $^{208}$Pb+$^{208}$Pb collisions. Recent measurements show that the inclusive $v_n$ exhibit modest differences ($<10$--20\%) between these two systems as a function of centrality, except in the central collisions where the difference for $v_2$ is significantly larger~\cite{Acharya:2018ihu,Sirunyan:2019wqp,Aad:2019xmh}. These are sensitive to the differences in the initial eccentricities and viscous effects in the two systems~\cite{Giacalone:2017dud,Eskola:2017bup}. Similarly, comparison of $v_n$ decorrelation between Xe+Xe and Pb+Pb, together with the comparison of inclusive $v_n$, could improve our understanding of the longitudinal structures of the QGP, and in particular answer the question whether the decorrelation is controlled by the overall system size or the shape of the overlap region.

The measurement is performed using the ATLAS inner detector (ID) and forward calorimeters (FCal) along with the trigger and data acquisition system~\cite{Aad:2008zzm,Aaboud:2016leb}. The ID measures charged particles over a pseudorapidity\footnote{ATLAS uses a right-handed coordinate system with its origin at the nominal interaction point (IP) in the center of the detector and the $z$-axis along the beam pipe. The $x$-axis points from the IP to the center of the LHC ring, and the $y$-axis points upward. Cylindrical coordinates $(r,\phi)$ are used in the transverse plane, $\phi$ being the azimuthal angle around the beam pipe. The pseudorapidity is defined in terms of the polar angle $\theta$ as $\eta=-\ln\tan(\theta/2)$.} range $|\eta|<2.5$ using a combination of silicon pixel detectors, silicon microstrip detectors, and a straw-tube transition radiation tracker, all immersed in a 2~T axial magnetic field~\cite{Aad:ID,ATLAS-TDR-19,PIX-2018-001}. The FCal measures the sum of the transverse energy $\sumET$ over $3.2<|\eta|<4.9$ to determine the event centrality, and uses copper and tungsten absorbers with liquid argon as the active medium.  The ATLAS trigger system~\cite{Aaboud:2016leb} consists of a level-1 (L1) trigger based on electronics, and a software-based high-level trigger.

This analysis uses 3~${\mathalpha{\mu}}{\textrm b}^{-1}$ of $\sqn=5.44~\TeV$ Xe+Xe data collected in 2017. The events are selected by requiring the total transverse energy deposited in the calorimeters over $|\eta|<4.9$ at L1 to be larger than 4~\GeV. In the offline
analysis, the $z$-position of the primary vertex~\cite{ATL-PHYS-PUB-2015-026} of each event is required to be within 100 mm of the IP. Events containing more than one inelastic interaction are suppressed by exploiting the correlation between the $\sumET$ measured in the FCal and the number of tracks associated with a primary vertex.  The event centrality classification is based on the $\sumET$ in the FCal~\cite{Aaboud:2018ves}. A Glauber model~\cite{Miller:2007ri,Aad:2011yr} is used to determine the mapping between $\sumET$ in the FCal and the centrality percentiles, as well as to estimate the average number of participating nucleons, $\Npart$, for each centrality interval.
 
Charged-particle tracks are reconstructed from ionization hits in the ID using a reconstruction procedure optimized for heavy-ion collisions~\cite{atlasreco}. Tracks used in this analysis are required to have $|\eta|<2.4$ and transverse momentum in the range $0.5<\pT<3~\GeV$. In addition, the point of closest approach of the track to the primary vertex is required to be within 1 mm in both the transverse and longitudinal directions. More details of the track selection can be found in Ref.~\cite{Aad:2019xmh}.
 
The efficiency $\epsilon(\pT,\eta)$ of the track reconstruction and track selection requirements is evaluated using minimum-bias Xe+Xe Monte Carlo (MC) events produced with the $\HIJING$~\cite{Gyulassy:1994ew} event generator with the effect of flow added via Ref.~\cite{Masera:2009zz}. The response of the detector was simulated~\cite{SOFT-2010-01} using $\GEANT$4~\cite{Agostinelli:2002hh}, and the resulting events are reconstructed with the same algorithms as applied to the data. The efficiency varies from 40\% to 73\% depending on $\eta$ and $\pT$, with an uncertainty of 1--4\% arising mainly from the uncertainty in the detector material budget. The rate of falsely reconstructed (fake) tracks, $f(\pT,\eta)$, is significant only for $\pT<0.8~\GeV$ in central collisions,  where it ranges from 2$\%$ for $\eta$ near zero to 6$\%$ for $|\eta|>2$.
 
The method and analysis procedure closely follow those established in Ref.~\cite{Aaboud:2017tql}, and are described briefly below. The $n^{\mathrm{th}}$-order azimuthal anisotropy in an event is estimated using the observed flow vectors:
\begin{linenomath}
\begin{align}
\bm{q}_n\equiv \Sigma_j w_j e^{\textrm{i} n\phi_j}/(\Sigma_j w_j)\;,
\end{align}
\end{linenomath}
where the sum runs over charged particles (for the ID) or calorimeter towers (for the FCal) in a specified $\eta$ interval, and $\phi_j$ and $w_j$ are the azimuthal angle and the weight assigned to each track or tower, respectively. The weight for the FCal is the $\eT$ of each tower, and the weight for the ID is calculated as $d(\eta,\phi)(1-f(\pT,\eta))/\epsilon(\pT,\eta)$ to correct for tracking performance~\cite{Aad:2014vba}. The additional factor $d(\eta,\phi)$, derived from the data, corrects for azimuthal nonuniformity of the detector performance in each $\eta$ interval.
 
The flow decorrelations are studied using product of flow vectors $\bm{q}_n(\eta)$ in the ID and $\bm{q}_n(\Eref)$ in the FCal~\cite{Khachatryan:2015oea} averaged over events in a given centrality interval,
\begin{linenomath}
\begin{align}\label{eq:2}
r_{n|n}(\eta)=\frac{\langle\bm{q}_n(-\eta)\bm{q}_n^{*}(\Eref)\rangle}{\langle\bm{q}_n(\eta)\bm{q}_n^{*}(\Eref)\rangle}
= \frac{\langle v_n(-\eta)v_n(\Eref)\,\cos\,n[\Phi_n(-\eta)-\Phi_n(\Eref)] \rangle}{\langle
v_n(\eta)v_n(\Eref)\,\cos\,n[\Phi_n(\eta)-\Phi_n(\Eref)] \rangle},
\end{align}
\end{linenomath}
where $\Eref$ is a reference pseudorapidity range in the FCal, common to both the numerator and the denominator. The $r_{n|n}$ correlator defined this way quantifies the decorrelation between $\eta$ and $-\eta$~\cite{Khachatryan:2015oea,Jia:2017kdq}. Three reference $\eta$ ranges, $3.2<|\Eref|<4.0$, $4.0<|\Eref|<4.9$ and $3.2<|\Eref|<4.9$ are used. Since $\lr{{\bm q}_n (-\eta) {\bm q}_n^{*}(\Eref)} = \lr{{\bm q}_n (\eta) {\bm q}_n^{*}(-\Eref)}$ for a symmetric system, the correlator is further symmetrized to enhance the statistics and reduce detector effects: $r_{n|n}(\eta) = \frac{\lr{{\bm q}_n (-\eta) {\bm q}_n^{*}(\Eref)+{\bm q}_n (\eta) {\bm q}_n^{*}(-\Eref)}}{\lr{{\bm q}_n (\eta){\bm q}_n^{*}(\Eref)+{\bm q}_n (-\eta){\bm q}_n^{*}(-\Eref)}}\;$.
 
If flow harmonics for two-particle correlation from two different $\eta$ factorize into single-particle harmonics, then it is expected that $r_{n|n}(\eta)=1$. Therefore, a value of $r_{n|n}(\eta)$ incompatible with unity implies a factorization-breaking effect due to longitudinal flow decorrelations. The deviation of $r_{n|n}$ from unity can be parameterized with a linear function, $r_{n|n}(\eta)= 1-2 F_n \eta$. The slope parameter $F_n$ is obtained via a simple linear regression of the $r_{n|n}(\eta)$ data~\cite{Aaboud:2017tql}. Using a Glauber model with a parameterized longitudinal structure, it was shown that $F_n\propto A_{\varepsilon_n} = \lr{\varepsilon_{n-}^{2}}/\lr{\varepsilon_{n}^{2}}$ with $\varepsilon_{n-}=|\mathcal{E}_{n-}|$~\cite{Behera:2020mol}, i.e.~$F_n$ is sensitive to the difference between the eccentricity for forward-going and backward-going participants. Since effects of viscosity partially cancels in the ratio, $F_n$ is less sensitive to late-time effects.
 
Systematic uncertainties in $r_{n|n}$ and the slope parameter $F_n$ arise from the uncertainties in the reconstruction and track selection efficiency, the acceptance reweighting procedure and the centrality definition. The systematic uncertainties are estimated by varying different aspects of the analysis, recalculating $r_{n|n}$ and $F_n$ and comparing them with the nominal values. The systematic uncertainty associated with fake tracks is estimated by loosening the requirements on the transverse and longitudinal impact parameters~\cite{Aad:2019xmh}; the resulting changes are 1--2\% for $F_2$, 1--4\% for $F_3$, and 1--9\% for $F_4$. The uncertainty associated with $\epsilon(\pT, \eta)$ is evaluated to be less than 1\% for $F_n$. The effect of reweighting is studied by setting $d(\eta,\phi)=1$ and repeating the analysis. The change is found to be 0.6--2\% for $F_2$ and $F_3$, and 2--7\% for $F_4$. The uncertainty due to the centrality definition is estimated by varying the mapping between $\sumET$ and centrality percentiles; the influence is 0.5--4\% for $F_2$ and $F_3$, and 0.5--8\% for $F_4$.
 
\begin{figure}[h!]
\centering
\includegraphics[width=0.8\linewidth]{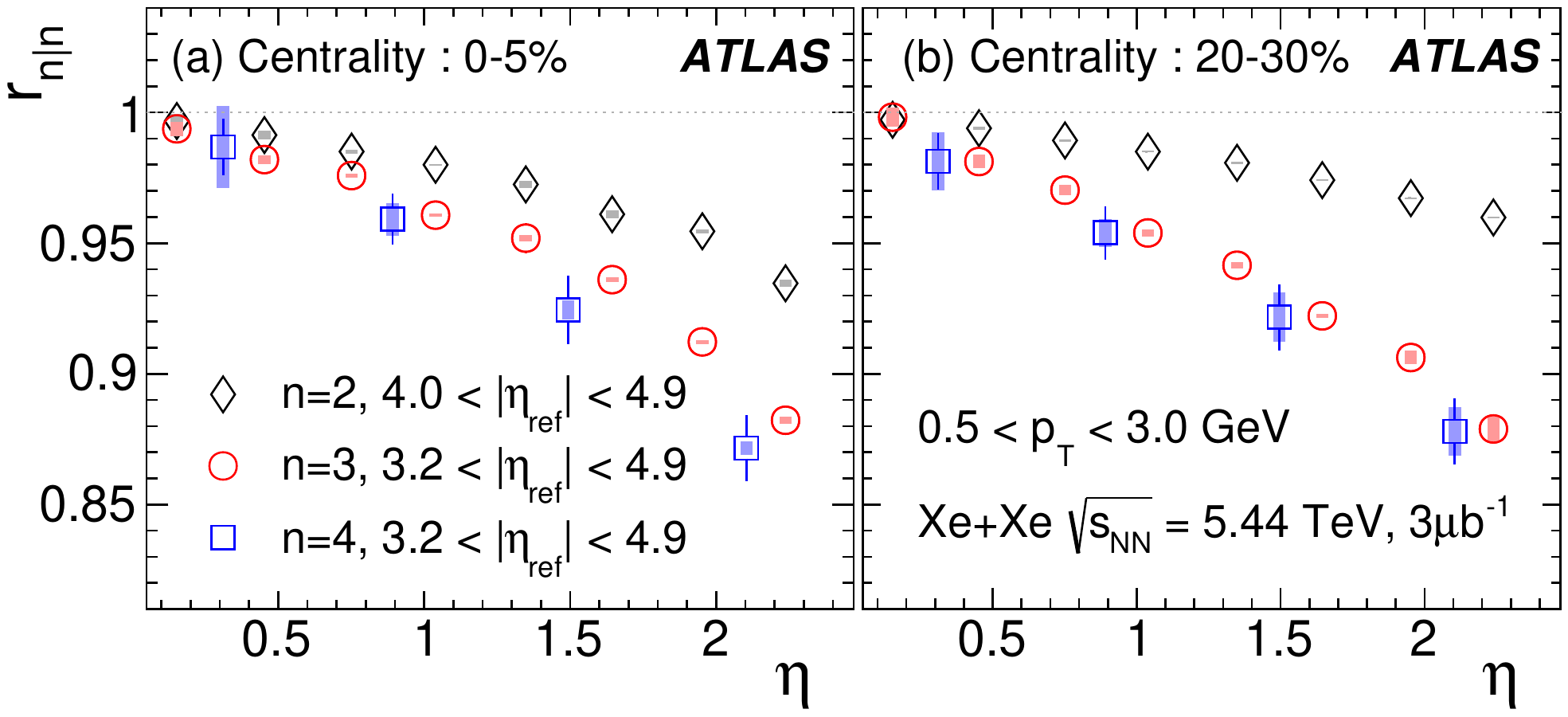}
\caption{The $\eta$ dependence of $r_{2|2}$, $r_{3|3}$ and $r_{4|4}$ in Xe+Xe collisions for two centrality intervals. The $|\eta_{\text{ref}}|$ is chosen to be $4.0<|\eta_{\text{ref}}|<4.9$ for $r_{2|2}$, and $3.2<|\eta_{\text{ref}}|<4.9$ for $r_{3|3}$ and $r_{4|4}$. The error bars and shaded boxes represent statistical and systematic uncertainties, respectively.}
\label{fig:1}
\end{figure}
Figure~\ref{fig:1} shows the measured $r_{n|n}(\eta)$ for $n=2$, 3 and 4 in two centrality intervals, quantifying the flow decorrelation between $\eta$ and $-\eta$ according to Eq.~\eqref{eq:2}. The $r_{n|n}$ values show an approximately linear decrease with $\eta$, implying stronger flow decorrelation at large $\eta$. The magnitudes of decorrelation for $r_{3|3}$ and $r_{4|4}$ are significantly larger than that for $r_{2|2}$. The range $4.0<|\Eref|<4.9$ chosen for $r_{2|2}$ is different from the range $3.2<|\Eref|<4.9$ used for $r_{3|3}$ and $r_{4|4}$ in order to reduce sensitivity to nonflow correlations; this is further discussed below.

\begin{figure}[h!]
\centering
\includegraphics[width=0.75\linewidth]{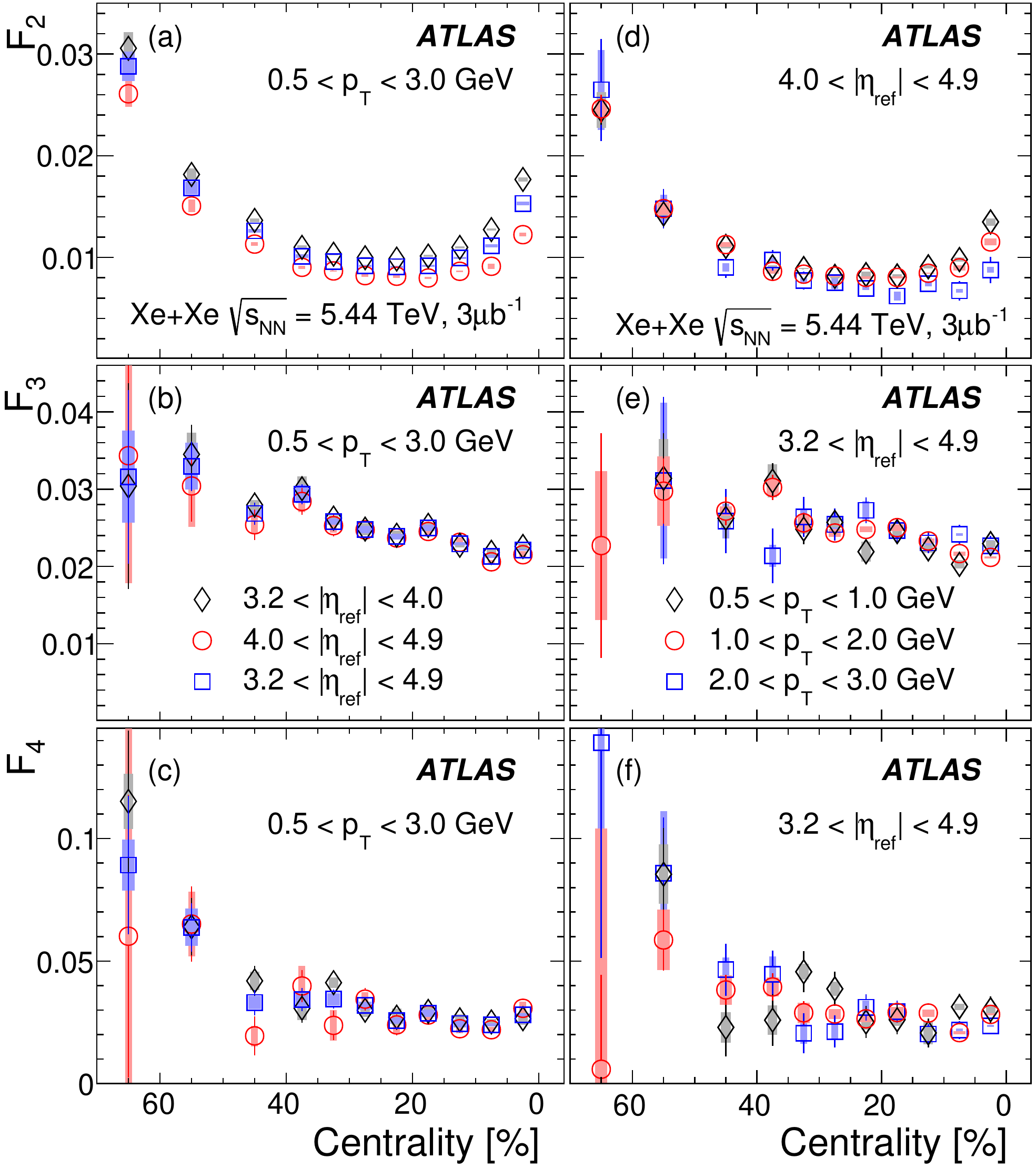}
\caption{The centrality dependence of $F_n$ calculated for three $|\eta_{\text{ref}}|$ ranges (left) and three $p_{\text{T}}$ ranges (right) for $n=2$ (top row), $n=3$ (middle row) and $n=4$ (bottom row). The error bars and shaded boxes represent statistical and systematic uncertainties, respectively.}
\label{fig:2}
\end{figure}

The slope parameters $F_n$ for $r_{n|n}$ are summarized in Figure~\ref{fig:2} as a function of centrality percentile with smaller percentile corresponds to more-central collisions. The left panels show the $F_n$ for three $|\Eref|$ ranges and right panels show the $F_n$ for three $\pT$ ranges. Within uncertainties, $F_3$ and $F_4$ show very weak dependence on centrality. The $F_2$ values, on the other hand, show a strong centrality dependence: they are smallest in the 20--30\% centrality interval and larger towards more-central or more-peripheral collisions. This strong centrality dependence is related to the fact that $v_2$ is dominated by the average elliptic geometry in mid-central collisions and therefore is less affected by decorrelations, while it is dominated by fluctuation-driven collision geometries in central and peripheral collisions~\cite{Bozek:2015bna,Pang:2015zrq}.
 
Figure~\ref{fig:2} also shows that $F_2$ has sizable variation between choices of $|\Eref|$ or $\pT$ in central and mid-central collisions. The contribution from nonflow correlations associated with back-to-back dijets are expected to contribute to the denominator more than the numerator due to a small gap between $\eta$ and $\Eref$, and therefore tend to increase the $F_n$ values~\cite{Aaboud:2017tql,Jia:2017hbm}. Such nonflow contributions are expected to be larger for smaller $|\Eref|$ or larger $\pT$. However, although the data show a larger $F_2$ for smaller $|\Eref|$ compatible with nonflow, they show a smaller $F_2$ for larger $\pT$, opposite to the expectation from nonflow contributions. Such $\pT$ and $\Eref$ dependences are most significant in ultra-central collisions, suggesting a nonlinear behavior of $v_2$ decorrelation due to disappearance of average elliptic geometry in these collisions. Within uncertainties, the $F_3$ and $F_4$, as well as the original $r_{3|3}$ and $r_{4|4}$, show no differences between various $\pT$ or $|\Eref|$ ranges, suggesting that they are not affected by nonflow. Based on results in Figure~\ref{fig:2},  $4.0<|\Eref|<4.9$ is chosen for $F_2$ to reduce nonflow, but a wider range $3.2<|\Eref|<4.9$ is chosen for $F_3$ and $F_4$ to improve the precision of the measurement.
 
\begin{figure}[h!]
\centering
\includegraphics[width=0.75\linewidth]{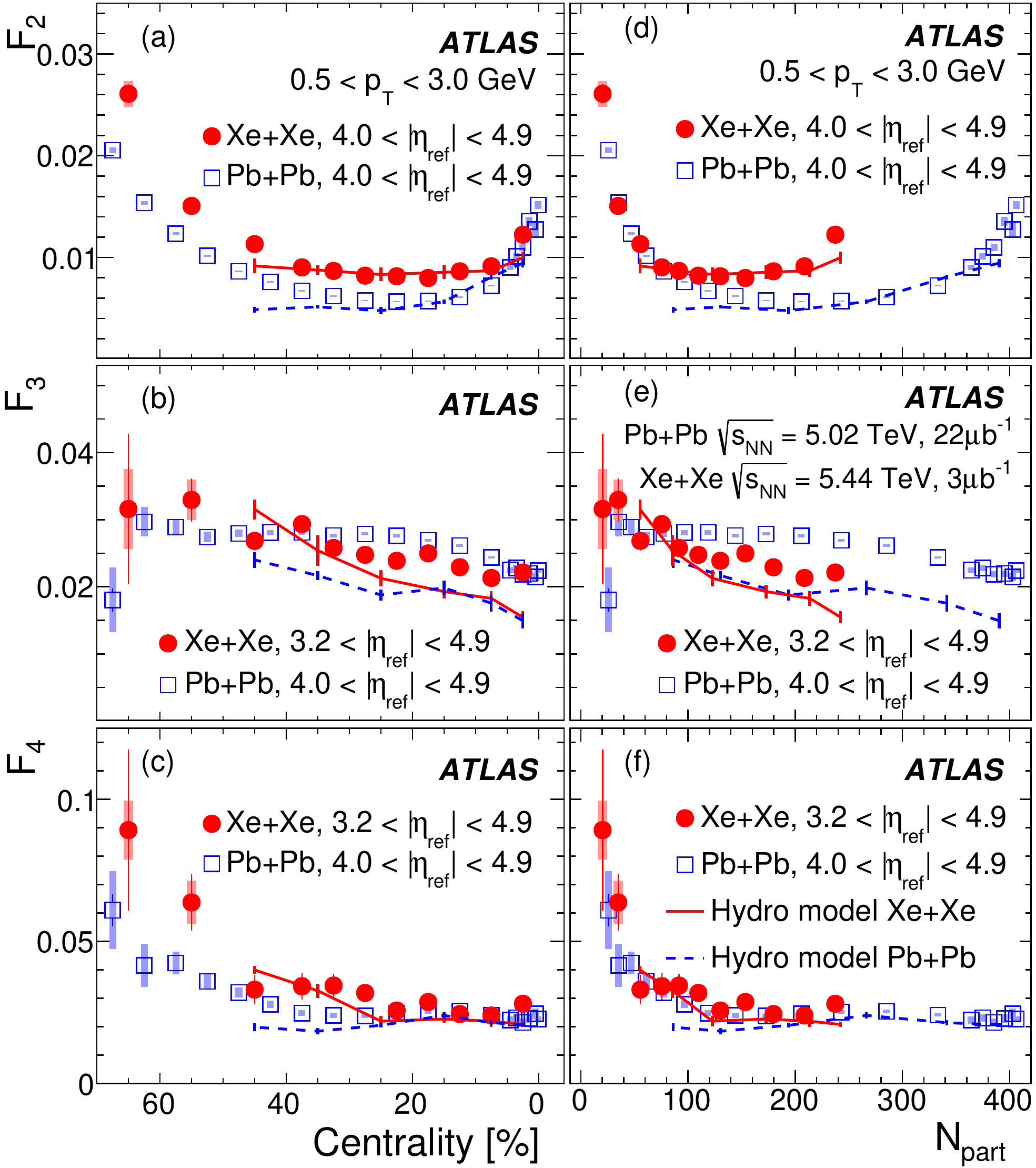}
\caption{ The $F_n$ compared between Xe+Xe and Pb+Pb~\cite{Aaboud:2017tql} collisions as a function of centrality percentiles (left) and $N_{\text{part}}$ (right) for $n=2$ (top row), $n=3$ (middle row) and $n=4$ (bottom row). The error bars and shaded boxes on the data represent statistical and systematic uncertainties, respectively. The results from a hydrodynamic model~\cite{Pang:2018zzo,Wu:2018cpc} are shown as solid lines (Xe+Xe) and dashed lines (Pb+Pb) with the vertical error bars denoting statistical uncertainty of the model predictions.}
\label{fig:3}
\end{figure}

To gain insights into the system-size dependence of the longitudinal fluctuations, Figure~\ref{fig:3} compares the $F_n$ from the Xe+Xe system with those obtained from the Pb+Pb system at $\sqn = 5.02~\TeV$ from Ref.~\cite{Aaboud:2017tql} as a function of centrality percentile (left column) or $\npart$ (right column). For both systems, $F_2$ shows a strong dependence on centrality percentile and $\npart$, while the $F_3$ and $F_4$ each show rather weak dependence. The $F_4$ values depend weakly on both centrality percentile and $\npart$, and they agree between the two systems. In the noncentral collisions (centrality percentiles $\gtrsim 30\%$ or $\npart\lesssim80$), the $F_2$ for the two systems agree only as a function of $\npart$, while the $F_3$ agree as a function of either centrality percentiles or $\npart$. In the mid-central collisions, $F_2$ is much larger in Xe+Xe collisions than in Pb+Pb collisions, while an opposite trend is observed for $F_3$. This reverse system-size ordering between $F_2$ and $F_3$ is also observed for $A_{\varepsilon_2}$ and $A_{\varepsilon_3}$ from Ref.~\cite{Behera:2020mol}, which strongly suggests that the flow decorrelations are driven by longitudinal fluctuations of the eccentricity vector in the initial state. The data are also compared with results from a hydrodynamic model with longitudinal fluctuations included~\cite{Pang:2018zzo,Wu:2018cpc}. The model quantitatively describes the behavior of $F_2$ and $F_4$ in mid-central collisions, but fails to describe the magnitude of $F_3$ and the splitting between the two systems, pointing to an inadequate description of the initial state and its system-size dependence implemented in this model.
 
\begin{figure}[t!]
\centering
\includegraphics[width=1.\linewidth]{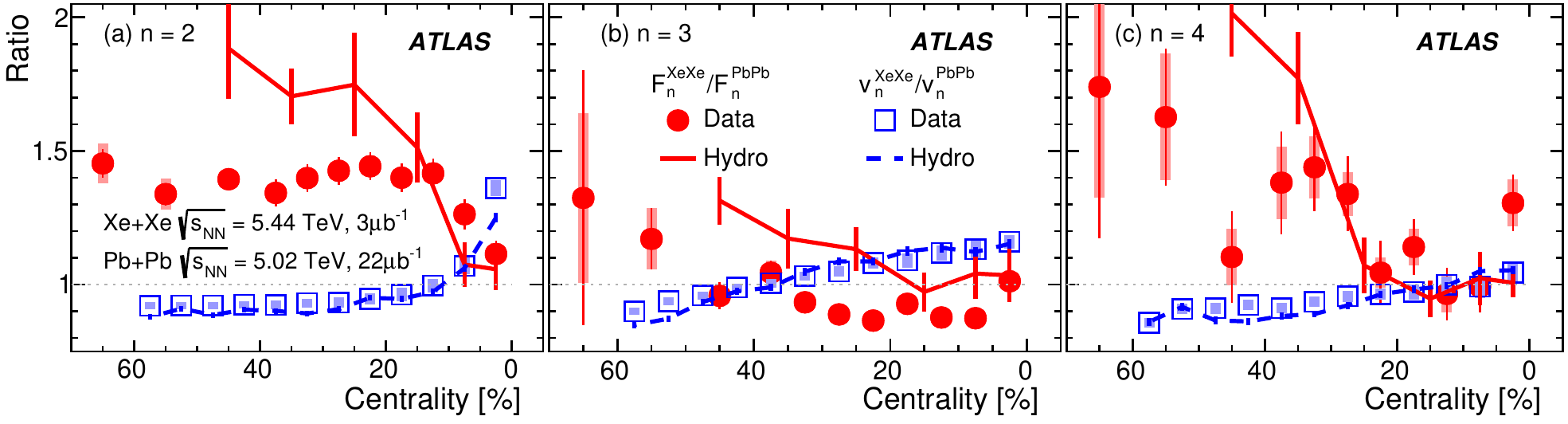}
\caption{The ratios $F_n^{\text{XeXe}}/F_n^{\text{PbPb}}$ from data~\cite{Aaboud:2017tql} (solid
symbols) and model~\cite{Pang:2018zzo,Wu:2018cpc} (solid lines) and $v_n^{\text{XeXe}}/v_n^{\text{PbPb}}$ from data~\cite{Aad:2019xmh} (open symbols) and model~\cite{Giacalone:2017dud} (dashed lines) as a function of centrality for $n=2$ (left), $n=3$ (middle panel) and $n=4$ (right), respectively. The error bars and shaded boxes on the data represent statistical and systematic uncertainties, respectively. The vertical error bars on the theory calculations represent the statistical uncertainties.}
\label{fig:4}
\end{figure}

To help further understand the relationship between the transverse harmonic flow and its longitudinal fluctuations, Figure~\ref{fig:4} compares the ratios of flow decorrelation $F_n^{\text{XeXe}}/F_n^{\text{PbPb}}$ ($F_n$-ratios) with ratios of flow harmonics $v_n^{\text{XeXe}}/v_n^{\text{PbPb}}$ ($v_n$-ratios) from Ref.~\cite{Aad:2019xmh} as a function of centrality percentile. While the $v_n$-ratios all decrease with centrality percentile, the $F_n$-ratios increase with centrality percentile; this opposite trend implies that when the ratio of average flow is larger, the ratio of its relative fluctuations in the longitudinal direction is smaller and vice versa. Beyond this overall opposite trend, there are other contrasting features between the two types of ratios. The $F_2$-ratio is always above one, while the $v_2$-ratio decreases to below one around 10--20\% centrality; the $F_2$-ratio is larger than the $v_2$-ratio except in the 0--5\% centrality interval, where the $v_2$-ratio is enhanced due to the deformation of the Xe nucleus~\cite{Giacalone:2017dud}. The differences between the $F_3$-ratio and the $v_3$-ratio are smaller, but with different centrality dependencies: while the $v_3$-ratio decreases nearly linearly with centrality percentile, the $F_3$-ratio first decreases and then increases as a function of centrality percentile. The $F_4$-ratio has larger uncertainties, but shows much stronger centrality dependence compared with the $v_4$-ratio.
 
Figure~\ref{fig:4} compares these ratios with hydrodynamic model calculations~\cite{Giacalone:2017dud,Pang:2018zzo,Wu:2018cpc}. The advantage of comparison in terms of ratios is that the model uncertainties in the initial-state geometry as well as final-state dynamics are expected to partially cancel out. While the calculations from Ref.~\cite{Giacalone:2017dud} quantitatively describe the trend of the $v_n$-ratios, they agree less well with the $F_n$-ratios and in particular the model~\cite{Pang:2018zzo,Wu:2018cpc} overestimates the $F_2$- and $F_3$-ratios for centrality percentiles beyond 20--30\%. Therefore, these hydrodynamic models fail to describe the longitudinal flow fluctuations and their system-size dependence trends, even though they have been tuned to describe the overall transverse collective dynamics. This failure is likely due to an inadequate description of the longitudinal structure of the initial state in these models. In fact, a recent calculation~\cite{Behera:2020mol} based on a simple Glauber model with the parameterized longitudinal structure was able to describe simultaneously the system-size dependence of the $v_n$ decorrelation and inclusive $v_n$, supporting this conjecture. One future direction is to develop a framework based on the three-dimensional initial condition dynamically generated from gluon saturation physics, coupled with a hydrodynamic model~\cite{Schenke:2016ksl,McDonald:2020oyf}. The part of ${\mathcal{E}}_{n-}$ arising from gluon saturation is related to the saturation scale ($Q_s$) controlled by the overall system size, while that arising from the forward-backward asymmetry is related to the shape of the overlap controlled by the centrality. Therefore, one could fix the $Q_s$ evolution in the Pb+Pb and make predictions in the Xe+Xe system, which will help to separate different initial state effects. The system-size dependence of the $v_n$ and $v_n$ decorrelation data provide important input to stimulate further theoretical efforts along this direction.
 
In summary, ATLAS presents the first measurement of longitudinal decorrelations for harmonic flow $v_n$ in Xe+Xe collisions at $\sqn=5.44~\TeV$, based on 3~${\mathalpha{\mu}}\textrm{b}^{-1}$ of data collected at the LHC. The $v_n$ decorrelations are nearly independent of centrality percentile and $\pT$ for $n=3$ and 4. For $n=2$, the $v_n$ decorrelations are smallest in mid-central collisions and increases for more-central or more-peripheral collisions, and also depends on $\pT$. A comparison with Pb+Pb collisions at $\sqn=5.02~\TeV$ shows that the $v_2$ decorrelation is larger in Xe+Xe collisions than in Pb+Pb collisions in most of the centrality range, while the opposite trend is observed for $v_3$ decorrelation. This reverse ordering is consistent with the expected behavior of eccentricity decorrelations in the two systems, and is not observed for the ratios of $v_2$ and $v_3$ between the two systems. Hydrodynamic models are found to describe the ratios of $v_n$ between Xe+Xe and Pb+Pb, but fail to describe most of the magnitudes and trends of the ratios of the $v_n$ decorrelations between Xe+Xe and Pb+Pb. This suggests that current models tuned to describe the transverse dynamics do not describe the longitudinal structure of the initial-state geometry.
 
Understanding the initial conditions and early-time effects is vital for adequate modeling of heavy-ion collisions~\cite{Busza:2018rrf}. System-size dependence of flow decorrelations, together with measurements of the inclusive flow harmonics, provide new insights and an important lever-arm to separate effects of the longitudinal structure of the initial state from other early-time and late-time effects. This measurement gives important input for complete modeling of the three-dimensional initial conditions and space-time dynamics of heavy-ion collisions used in hydrodynamic models.
 
\section*{Acknowledgments}

We thank CERN for the very successful operation of the LHC, as well as the
support staff from our institutions without whom ATLAS could not be
operated efficiently.
 
We acknowledge the support of ANPCyT, Argentina; YerPhI, Armenia; ARC, Australia; BMWFW and FWF, Austria; ANAS, Azerbaijan; SSTC, Belarus; CNPq and FAPESP, Brazil; NSERC, NRC and CFI, Canada; CERN; ANID, Chile; CAS, MOST and NSFC, China; Minciencias, Colombia; MSMT CR, MPO CR and VSC CR, Czech Republic; DNRF and DNSRC, Denmark; IN2P3-CNRS and CEA-DRF/IRFU, France; SRNSFG, Georgia; BMBF, HGF and MPG, Germany; GSRT, Greece; RGC and Hong Kong SAR, China; ISF and Benoziyo Center, Israel; INFN, Italy; MEXT and JSPS, Japan; CNRST, Morocco; NWO, Netherlands; RCN, Norway; MNiSW and NCN, Poland; FCT, Portugal; MNE/IFA, Romania; JINR; MES of Russia and NRC KI, Russian Federation; MESTD, Serbia; MSSR, Slovakia; ARRS and MIZ\v{S}, Slovenia; DST/NRF, South Africa; MICINN, Spain; SRC and Wallenberg Foundation, Sweden; SERI, SNSF and Cantons of Bern and Geneva, Switzerland; MOST, Taiwan; TAEK, Turkey; STFC, United Kingdom; DOE and NSF, United States of America. In addition, individual groups and members have received support from BCKDF, CANARIE, Compute Canada, CRC and IVADO, Canada; Beijing Municipal Science \& Technology Commission, China; COST, ERC, ERDF, Horizon 2020 and Marie Sk{\l}odowska-Curie Actions, European Union; Investissements d'Avenir Labex, Investissements d'Avenir Idex and ANR, France; DFG and AvH Foundation, Germany; Herakleitos, Thales and Aristeia programmes co-financed by EU-ESF and the Greek NSRF, Greece; BSF-NSF and GIF, Israel; La Caixa Banking Foundation, CERCA Programme Generalitat de Catalunya and PROMETEO and GenT Programmes Generalitat Valenciana, Spain; G\"{o}ran Gustafssons Stiftelse, Sweden; The Royal Society and Leverhulme Trust, United Kingdom.
 
The crucial computing support from all WLCG partners is acknowledged gratefully, in particular from CERN, the ATLAS Tier-1 facilities at TRIUMF (Canada), NDGF (Denmark, Norway, Sweden), CC-IN2P3 (France), KIT/GridKA (Germany), INFN-CNAF (Italy), NL-T1 (Netherlands), PIC (Spain), ASGC (Taiwan), RAL (UK) and BNL (USA), the Tier-2 facilities worldwide and large non-WLCG resource providers. Major contributors of computing resources are listed in Ref.~\cite{ATL-SOFT-PUB-2020-001}.
 

\printbibliography

@article{Busza:2018rrf,
    author = "Busza, Wit and Rajagopal, Krishna and van der Schee, Wilke",
    title = "{Heavy Ion Collisions: The Big Picture, and the Big Questions}",
    eprint = "1802.04801",
    archivePrefix = "arXiv",
    primaryClass = "hep-ph",
    reportNumber = "MIT-CTP-4892, MIT-CTP/4892",
    doi = "10.1146/annurev-nucl-101917-020852",
    journal = "Ann. Rev. Nucl. Part. Sci.",
    volume = "68",
    pages = "339--376",
    year = "2018"
}

@article{Behera:2020mol,
    author = "Behera, Arabinda and Nie, Maowu and Jia, Jiangyong",
    title = "{Longitudinal eccentricity decorrelations in heavy ion collisions}",
    eprint = "2003.04340",
    archivePrefix = "arXiv",
    primaryClass = "nucl-th",
    doi = "10.1103/PhysRevResearch.2.023362",
    journal = "Phys. Rev. Res.",
    volume = "2",
    number = "2",
    pages = "023362",
    year = "2020"
}

@article{Schenke:2016ksl,
    author = "Schenke, Bjoern and Schlichting, Soeren",
    title = "{3D glasma initial state for relativistic heavy ion collisions}",
    eprint = "1605.07158",
    archivePrefix = "arXiv",
    primaryClass = "hep-ph",
    doi = "10.1103/PhysRevC.94.044907",
    journal = "Phys. Rev. C",
    volume = "94",
    number = "4",
    pages = "044907",
    year = "2016"
}

@inproceedings{McDonald:2020oyf,
    author = "McDonald, Scott and Jeon, Sangyong and Gale, Charles",
    title = "{Exploring Longitudinal Observables with 3+1D IP-Glasma}",
    booktitle = "{28th International Conference on Ultrarelativistic Nucleus-Nucleus Collisions}",
    eprint = "2001.08636",
    archivePrefix = "arXiv",
    primaryClass = "nucl-th",
    month = "1",
    year = "2020"
}

@article{Sakai:2020pjw,
    author = "Sakai, Azumi and Murase, Koichi and Hirano, Tetsufumi",
    title = "{Rapidity decorrelation of anisotropic flow caused by hydrodynamic fluctuations}",
    eprint = "2003.13496",
    archivePrefix = "arXiv",
    primaryClass = "nucl-th",
    month = "3",
    year = "2020"
}

@article{ALICE:2016kpq,
			author         = "{ALICE Collaboration}",
			title          = "{Correlated Event-by-Event Fluctuations of Flow Harmonics
												in Pb-Pb Collisions at $\sqrt{s_{\textrm{NN}}}=2.76$ TeV}",
			xcollaboration  = "ALICE",
			journal        = "Phys. Rev. Lett.",
			volume         = "117",
			year           = "2016",
			pages          = "182301",
			doi            = "10.1103/PhysRevLett.117.182301",
			eprint         = "1604.07663",
			archivePrefix  = "arXiv",
			primaryClass   = "nucl-ex",
			reportNumber   = "CERN-EP-2016-102",
			SLACcitation   = "%%CITATION = ARXIV:1604.07663;%%"
}

@article{Sirunyan:2017fts,
			author         = "{CMS Collaboration}",
			title          = "{Non-Gaussian elliptic-flow fluctuations in PbPb
												collisions at $\sqrt{\smash[b]{s_{_\text{NN}}}} = 5.02$
												TeV}",
			xcollaboration  = "CMS",
			journal        = "Phys. Lett. B",
			volume         = "789",
			year           = "2019",
			pages          = "643-665",
			doi            = "10.1016/j.physletb.2018.11.063",
			eprint         = "1711.05594",
			archivePrefix  = "arXiv",
			primaryClass   = "nucl-ex",
			reportNumber   = "CMS-HIN-16-019, CERN-EP-2017-268",
			SLACcitation   = "%%CITATION = ARXIV:1711.05594;%%"
}

@article{Adam:2019woz,
			author         = "{STAR Collaboration}",
			title          = "{Azimuthal Harmonics in Small and Large Collision Systems
												at RHIC Top Energies}",
			collaboration  = "STAR",
			journal        = "Phys. Rev. Lett.",
			volume         = "122",
			year           = "2019",
			number         = "17",
			pages          = "172301",
			doi            = "10.1103/PhysRevLett.122.172301",
			eprint         = "1901.08155",
			archivePrefix  = "arXiv",
			primaryClass   = "nucl-ex",
			SLACcitation   = "%%CITATION = ARXIV:1901.08155;%%"
}

@article{Acharya:2018lmh,
			author         = "{ALICE Collaboration}",
			title          = "{Energy dependence and fluctuations of anisotropic flow
												in Pb-Pb collisions at $ \sqrt{s_{\mathrm{NN}}}=5.02 $ and
												2.76 TeV}",
			xcollaboration  = "ALICE",
			journal        = "JHEP",
			volume         = "07",
			year           = "2018",
			pages          = "103",
			doi            = "10.1007/JHEP07(2018)103",
			eprint         = "1804.02944",
			archivePrefix  = "arXiv",
			primaryClass   = "nucl-ex",
			reportNumber   = "CERN-EP-2018-057",
			SLACcitation   = "%%CITATION = ARXIV:1804.02944;%%"
}

@article{Aaboud:2018ves,
			author         = "{ATLAS Collaboration}",
			title          = "{Measurement of the azimuthal anisotropy of charged
												particles produced in $\sqrt{s_{_\text {NN}}}$ = 5.02 TeV
												Pb+Pb collisions with the ATLAS detector}",
			xcollaboration  = "ATLAS",
			journal        = "Eur. Phys. J. C",
			volume         = "78",
			year           = "2018",
			number         = "12",
			pages          = "997",
			doi            = "10.1140/epjc/s10052-018-6468-7",
			eprint         = "1808.03951",
			archivePrefix  = "arXiv",
			primaryClass   = "nucl-ex",
			reportNumber   = "CERN-EP-2018-194",
			SLACcitation   = "%%CITATION = ARXIV:1808.03951;%%"
}

@article{Acharya:2018ihu,
			author         = "{ALICE Collaboration}",
			title          = "{Anisotropic flow in Xe-Xe collisions at
												$\sqrt{s_{\textrm{NN}}} = 5.44$ TeV}",
			xcollaboration  = "ALICE",
			journal        = "Phys. Lett. B",
			volume         = "784",
			year           = "2018",
			pages          = "82-95",
			doi            = "10.1016/j.physletb.2018.06.059",
			eprint         = "1805.01832",
			archivePrefix  = "arXiv",
			primaryClass   = "nucl-ex",
			reportNumber   = "CERN-EP-2018-090",
			SLACcitation   = "%%CITATION = ARXIV:1805.01832;%%"
}

@article{Aad:2019xmh,
			author         = "{ATLAS Collaboration}",
			title          = "{Measurement of the azimuthal anisotropy of
												charged-particle production in Xe+Xe collisions at
												$\sqrt{s_{\mathrm{NN}}}=5.44$ TeV with the ATLAS
												detector}",
			xcollaboration  = "ATLAS",
    doi = "10.1103/PhysRevC.101.024906",
    journal = "Phys. Rev. C",
    volume = "101",
    number = "2",
    pages = "024906",
    year = "2020",
			eprint         = "1911.04812",
			archivePrefix  = "arXiv",
			primaryClass   = "nucl-ex",
			reportNumber   = "CERN-EP-2019-227",
			SLACcitation   = "%%CITATION = ARXIV:1911.04812;%%"
}

@article{Giacalone:2017dud,
			author         = "Giacalone, Giuliano and Noronha-Hostler, Jacquelyn and
												Luzum, Matthew and Ollitrault, Jean-Yves",
			title          = "{Hydrodynamic predictions for 5.44 TeV Xe+Xe collisions}",
			journal        = "Phys. Rev. C",
			volume         = "97",
			year           = "2018",
			number         = "3",
			pages          = "034904",
			doi            = "10.1103/PhysRevC.97.034904",
			eprint         = "1711.08499",
			archivePrefix  = "arXiv",
			primaryClass   = "nucl-th",
			reportNumber   = "SACLAY-T17-179",
			SLACcitation   = "%%CITATION = ARXIV:1711.08499;%%"
}

@article{Eskola:2017bup,
			author         = "Eskola, K. J. and Niemi, H. and Paatelainen, R. and
												Tuominen, K.",
			title          = "{Predictions for multiplicities and flow harmonics in
												5.44 TeV Xe+Xe collisions at the CERN Large Hadron
												Collider}",
			journal        = "Phys. Rev. C",
			volume         = "97",
			year           = "2018",
			number         = "3",
			pages          = "034911",
			doi            = "10.1103/PhysRevC.97.034911",
			eprint         = "1711.09803",
			archivePrefix  = "arXiv",
			primaryClass   = "hep-ph",
			SLACcitation   = "%%CITATION = ARXIV:1711.09803;%%"
}

@article{Sirunyan:2019wqp,
			author         = "{CMS Collaboration}",
			title          = "{Charged-particle angular correlations in XeXe collisions
												at $\sqrt{s_{_\mathrm{NN}}}=$ 5.44 TeV}",
			xcollaboration  = "CMS",
			journal        = "Phys. Rev. C",
			volume         = "100",
			year           = "2019",
			number         = "4",
			pages          = "044902",
			doi            = "10.1103/PhysRevC.100.044902",
			eprint         = "1901.07997",
			archivePrefix  = "arXiv",
			primaryClass   = "hep-ex",
			reportNumber   = "CMS-HIN-18-001, CERN-EP-2018-345",
			SLACcitation   = "%%CITATION = ARXIV:1901.07997;%%"
}

@Article{SOFT-2010-01,
		author         = "{ATLAS Collaboration}",
		title          = "{The ATLAS Simulation Infrastructure}",
		journal        = "Eur. Phys. J. C",
		volume         = "70",
		year           = "2010",
		pages          = "823",
		doi            = "10.1140/epjc/s10052-010-1429-9",
		eprint         = "1005.4568",
		archivePrefix  = "arXiv",
		primaryClass   = "physics.ins-det",
}

@Booklet{ATL-PHYS-PUB-2015-026,
			author       = "{ATLAS Collaboration}",
			title        = "{Vertex Reconstruction Performance of the ATLAS Detector
											 at $\sqrt{s} = 13~\TeV$}",
			howpublished = "ATL-PHYS-PUB-2015-026",
			year         = "2015",
			%reportNumber  = "ATL-PHYS-PUB-2015-026",
			url           = "https://cds.cern.ch/record/2037717",
}

@Booklet{ATLAS-TDR-19,
		 author       = "{ATLAS Collaboration}",
		 title        = "ATLAS Insertable B-Layer Technical Design Report",
		 howpublished = "{ATLAS-TDR-19, 2010,
{\scriptsize{URL}}:\url{https://cds.cern.ch/record/1291633}}",
		 %year         = "2010",
		 %url        = "https://cds.cern.ch/record/1291633",
		 note     = "{\textit{ATLAS Insertable B-Layer Technical Design Report Addendum}},
ATLAS-TDR-19-ADD-1, 2012, {\scriptsize{URL}}: \url{https://cds.cern.ch/record/1451888}",
}

@Article{PIX-2018-001,
		author         = "Abbott, B. and others",
		title          = "{Production and integration of the ATLAS Insertable B-Layer}",
		journal        = "JINST",
		volume         = "13",
		year           = "2018",
		pages          = "T05008",
		doi            = "10.1088/1748-0221/13/05/T05008",
		eprint         = "1803.00844",
		archivePrefix  = "arXiv",
		primaryClass   = "physics.ins-det",
}

@article{atlasreco,
			author         = "{ATLAS Collaboration}",
			title         = "{The Optimization of ATLAS Track Reconstruction in Dense
											 Environments}",
			journal        = "{ATL-PHYS-PUB-2015-006}",
		year         = "2015",
			url           = "https://cds.cern.ch/record/2002609"
}

@article{Wu:2018cpc,
			author         = "Wu, Xiang-Yu and Pang, Long-Gang and Qin, Guang-You and
												Wang, Xin-Nian",
			title          = "{Longitudinal fluctuations and decorrelations of
												anisotropic flows at energies available at the CERN Large
												Hadron Collider and at the BNL Relativistic Heavy Ion
												Collider}",
			journal        = "Phys. Rev. C",
			volume         = "98",
			year           = "2018",
			number         = "2",
			pages          = "024913",
			doi            = "10.1103/PhysRevC.98.024913",
			eprint         = "1805.03762",
			archivePrefix  = "arXiv",
			primaryClass   = "nucl-th",
			SLACcitation   = "%%CITATION = ARXIV:1805.03762;%%"
}

@article{Pang:2018zzo,
			author         = "Pang, Long-Gang and Petersen, Hannah and Wang, Xin-Nian",
			title          = "{Pseudorapidity distribution and decorrelation of
												anisotropic flow within the open-computing-language
												implementation CLVisc hydrodynamics}",
			journal        = "Phys. Rev. C",
			volume         = "97",
			year           = "2018",
			number         = "6",
			pages          = "064918",
			doi            = "10.1103/PhysRevC.97.064918",
			eprint         = "1802.04449",
			archivePrefix  = "arXiv",
			primaryClass   = "nucl-th",
			SLACcitation   = "%%CITATION = ARXIV:1802.04449;%%"
}

@article{Pang:2015zrq,
			author         = "Pang, Long-Gang and Petersen, Hannah and Qin, Guang-You
												and Roy, Victor and Wang, Xin-Nian",
			title          = "{Decorrelation of anisotropic flow along the longitudinal
												direction}",
			journal        = "Eur. Phys. J. A",
			volume         = "52",
			year           = "2016",
			pages          = "97",
			doi            = "10.1140/epja/i2016-16097-x",
			eprint         = "1511.04131",
			archivePrefix  = "arXiv",
			primaryClass   = "nucl-th",
			SLACcitation   = "%%CITATION = ARXIV:1511.04131;%%"
}

@article{Bozek:2015bna,
			author         = "Bozek, Piotr and Broniowski, Wojciech",
			title          = "{The torque effect and fluctuations of entropy deposition
												in rapidity in ultra-relativistic nuclear collisions}",
			journal        = "Phys. Lett. B",
			volume         = "752",
			year           = "2016",
			pages          = "206-211",
			doi            = "10.1016/j.physletb.2015.11.054",
			eprint         = "1506.02817",
			archivePrefix  = "arXiv",
			primaryClass   = "nucl-th",
			SLACcitation   = "%%CITATION = ARXIV:1506.02817;%%"
}

@article{Shen:2017bsr,
			author         = "Shen, Chun and Schenke, Björn",
			title          = "{Dynamical initial state model for relativistic heavy-ion
												collisions}",
			journal        = "Phys. Rev. C",
			volume         = "97",
			year           = "2018",
			number         = "2",
			pages          = "024907",
			doi            = "10.1103/PhysRevC.97.024907",
			eprint         = "1710.00881",
			archivePrefix  = "arXiv",
			primaryClass   = "nucl-th",
			SLACcitation   = "%%CITATION = ARXIV:1710.00881;%%"
}

@article{Bozek:2017qir,
			author         = "Bozek, Piotr and Broniowski, Wojciech",
			title          = "{Longitudinal decorrelation measures of flow magnitude
												and event-plane angles in ultrarelativistic nuclear
												collisions}",
			journal        = "Phys. Rev. C",
			volume         = "97",
			year           = "2018",
			number         = "3",
			pages          = "034913",
			doi            = "10.1103/PhysRevC.97.034913",
			eprint         = "1711.03325",
			archivePrefix  = "arXiv",
			primaryClass   = "nucl-th",
			SLACcitation   = "%%CITATION = ARXIV:1711.03325;%%"
}

@article{Aaboud:2017tql,
	author         = "{ATLAS Collaboration}",
	title          = "{Measurement of longitudinal flow decorrelations in Pb+Pb
	collisions at $\sqrt{s_{\text{NN}}}=2.76$ and 5.02 TeV
	with the ATLAS detector}",
	xcollaboration  = "ATLAS",
	journal        = "Eur. Phys. J. C",
	volume         = "78",
	year           = "2018",
	number         = "2",
	pages          = "142",
	doi            = "10.1140/epjc/s10052-018-5605-7",
	eprint         = "1709.02301",
	archivePrefix  = "arXiv",
	primaryClass   = "nucl-ex",
	reportNumber   = "CERN-EP-2017-191",
	SLACcitation   = "%%CITATION = ARXIV:1709.02301;%%"
}

@article{Khachatryan:2015oea,
			author         = "{CMS Collaboration}",
			title          = "{Evidence for transverse momentum and pseudorapidity
												dependent event plane fluctuations in PbPb and pPb
												collisions}",
			xcollaboration  = "CMS",
			journal        = "Phys. Rev. C",
			volume         = "92",
			year           = "2015",
			number         = "3",
			pages          = "034911",
			doi            = "10.1103/PhysRevC.92.034911",
			eprint         = "1503.01692",
			archivePrefix  = "arXiv",
			primaryClass   = "nucl-ex",
			reportNumber   = "CMS-HIN-14-012, CERN-PH-EP-2015-039",
			SLACcitation   = "%%CITATION = ARXIV:1503.01692;%%"
}

@article{Gale:2013da,
			author         = "Gale, Charles and Jeon, Sangyong and Schenke, Bjoern",
			title          = "{Hydrodynamic Modeling of Heavy-Ion Collisions}",
			journal        = "Int.~J.~Mod.~Phys.~A",
			volume         = "28",
			pages          = "1340011",
			doi            = "10.1142/S0217751X13400113",
			year           = "2013",
			eprint         = "1301.5893",
			archivePrefix  = "arXiv",
			primaryClass   = "nucl-th",
			xSLACcitation   = "%%CITATION = ARXIV:1301.5893;%%",
}

@article{Heinz:2013th,
			author         = "Heinz, Ulrich and Snellings, Raimond",
			title          = "{Collective Flow and Viscosity in Relativistic Heavy-Ion
												Collisions}",
			journal        = "Ann.~Rev.~Nucl.~Part.~Sci.",
			volume         = "63",
			pages          = "123",
			doi            = "10.1146/annurev-nucl-102212-170540",
			year           = "2013",
			eprint         = "1301.2826",
			archivePrefix  = "arXiv",
			primaryClass   = "nucl-th",
			xSLACcitation   = "%%CITATION = ARXIV:1301.2826;%%",
}

@article{Adare:2011tg,
			author         = "{PHENIX Collaboration}",
			title          = "{Measurements of Higher-Order Flow Harmonics in Au+Au
												Collisions at $\sqrt{s_{NN}} = 200$ GeV}",
			xcollaboration  = "PHENIX",
			journal        = "Phys.~Rev.~Lett.",
			volume         = "107",
			pages          = "252301",
			doi            = "10.1103/PhysRevLett.107.252301",
			year           = "2011",
			eprint         = "1105.3928",
			archivePrefix  = "arXiv",
			primaryClass   = "nucl-ex",
			xSLACcitation   = "%%CITATION = ARXIV:1105.3928;%%",
}

@article{ALICE:2011ab,
			author         = "{ALICE Collaboration}",
			title          = "{Higher Harmonic Anisotropic Flow Measurements of Charged
												Particles in Pb-Pb Collisions at $\sqrt{s_{NN}}=2.76$
												TeV}",
			xcollaboration  = "ALICE",
			journal        = "Phys.~Rev.~Lett.",
			volume         = "107",
			pages          = "032301",
			doi            = "10.1103/PhysRevLett.107.032301",
			year           = "2011",
			eprint         = "1105.3865",
			archivePrefix  = "arXiv",
			primaryClass   = "nucl-ex",
			xreportNumber   = "CERN-PH-EP-2011-073",
			xSLACcitation   = "%%CITATION = ARXIV:1105.3865;%%",
}

@article{Aad:2012bu,
			author         = "{ATLAS Collaboration}",
			title          = "{Measurement of the azimuthal anisotropy for charged
												particle production in $\sqrt{s_{_{NN}}}=2.76$ TeV lead-lead
												collisions with the ATLAS detector}",
			xcollaboration  = "ATLAS",
			journal        = "Phys.~Rev.~C",
			volume         = "86",
			pages          = "014907",
			doi            = "10.1103/PhysRevC.86.014907",
			year           = "2012",
			eprint         = "1203.3087",
			archivePrefix  = "arXiv",
			primaryClass   = "hep-ex",
			xreportNumber   = "CERN-PH-EP-2012-035",
			xSLACcitation   = "%%CITATION = ARXIV:1203.3087;%%",
}

@article{Chatrchyan:2013kba,
			author         = "{CMS Collaboration}",
			title          = "{Measurement of higher-order harmonic azimuthal
												anisotropy in PbPb collisions at $\sqrt{s_{NN}}$ = 2.76
												TeV}",
			xcollaboration  = "CMS",
			journal        = "Phys.~Rev.~C",
			volume         = "89",
			pages          = "044906",
			doi            = "10.1103/PhysRevC.89.044906",
			year           = "2014",
			eprint         = "1310.8651",
			archivePrefix  = "arXiv",
			primaryClass   = "nucl-ex",
			xreportNumber   = "CMS-HIN-11-005, CERN-PH-EP-2013-196",
			xSLACcitation   = "%%CITATION = ARXIV:1310.8651;%%",
}

@article{Aad:2013xma,
			author         = "{ATLAS Collaboration}",
			title          = "{Measurement of the distributions of event-by-event flow
												harmonics in lead-lead collisions at $\sqrt{s_{\mathrm{NN}}}=2.76~\TeV$ with
the
												ATLAS detector at the LHC}",
			xcollaboration  = "ATLAS",
			journal        = "JHEP",
			volume         = "11",
			pages          = "183",
			doi            = "10.1007/JHEP11(2013)183",
			year           = "2013",
			eprint         = "1305.2942",
			archivePrefix  = "arXiv",
			primaryClass   = "hep-ex",
			xreportNumber   = "CERN-PH-EP-2013-04, CERN-PH-EP-2013-04",
			xSLACcitation   = "%%CITATION = ARXIV:1305.2942;%%",
}

@article{Aad:2014fla,
			author         = "{ATLAS Collaboration}",
			title          = "{Measurement of event-plane correlations in
												$\sqrt{s_{NN}}=2.76$ TeV lead-lead collisions with the
												ATLAS detector}",
			xcollaboration  = "ATLAS",
			journal        = "Phys.~Rev.~C",
			volume         = "90",
			pages          = "024905",
			doi            = "10.1103/PhysRevC.90.024905",
			year           = "2014",
			eprint         = "1403.0489",
			archivePrefix  = "arXiv",
			primaryClass   = "hep-ex",
			xreportNumber   = "CERN-PH-EP-2014-021",
			xSLACcitation   = "%%CITATION = ARXIV:1403.0489;%%",
}

@article{Aad:2015lwa,
			author         = "{ATLAS Collaboration}",
			title          = "{Measurement of the correlation between flow harmonics of
												different order in lead-lead collisions at
												$\sqrt{s_{NN}}$=2.76 TeV with the ATLAS detector}",
			xcollaboration  = "ATLAS",
			journal        = "Phys. Rev. C",
			volume         = "92",
			year           = "2015",
			pages          = "034903",
			doi            = "10.1103/PhysRevC.92.034903",
			eprint         = "1504.01289",
			archivePrefix  = "arXiv",
			primaryClass   = "hep-ex",
			reportNumber   = "CERN-PH-EP-2015-079",
			SLACcitation   = "%%CITATION = ARXIV:1504.01289;%%"
}

@article{Luzum:2012wu,
			author         = "Luzum, Matthew and Ollitrault, Jean-Yves",
			title          = "{Extracting the shear viscosity of the quark-gluon plasma
												from flow in ultra-central heavy-ion collisions}",
			journal        = "Nucl.~Phys.~A",
			volume         = "904--905",
			pages          = "377--380",
			doi            = "10.1016/j.nuclphysa.2013.02.028",
			year           = "2013",
			eprint         = "1210.6010",
			archivePrefix  = "arXiv",
			primaryClass   = "nucl-th",
			xSLACcitation   = "%%CITATION = ARXIV:1210.6010;%%",
}

@article{Teaney:2010vd,
			author         = "Teaney, Derek and Yan, Li",
			title          = "{Triangularity and dipole asymmetry in relativistic heavy ion
												collisions}",
			journal        = "Phys.~Rev.~C",
			volume         = "83",
			pages          = "064904",
			doi            = "10.1103/PhysRevC.83.064904",
			year           = "2011",
			eprint         = "1010.1876",
			archivePrefix  = "arXiv",
			primaryClass   = "nucl-th",
			xSLACcitation   = "%%CITATION = ARXIV:1010.1876;%%",
}

@article{Gale:2012rq,
			author         = "Gale, Charles and Jeon, Sangyong and Schenke, Bjorn and
												Tribedy, Prithwish and Venugopalan, Raju",
			title          = "{Event-by-event Anisotropic Flow in Heavy-ion Collisions
												from Combined Yang-Mills and Viscous Fluid Dynamics}",
			journal        = "Phys.~Rev.~Lett.",
			volume         = "110",
			pages          = "012302",
			doi            = "10.1103/PhysRevLett.110.012302",
			year           = "2013",
			eprint         = "1209.6330",
			archivePrefix  = "arXiv",
			primaryClass   = "nucl-th",
			xSLACcitation   = "%%CITATION = ARXIV:1209.6330;%%",
}

@article{Niemi:2012aj,
			author         = "Niemi, H. and Denicol, G. S. and Holopainen, H. and
												Huovinen, P.",
			title          = "{Event-by-event distributions of azimuthal asymmetries in
												ultrarelativistic heavy-ion collisions}",
			journal        = "Phys.~Rev.~C",
			volume         = "87",
			pages          = "054901",
			doi            = "10.1103/PhysRevC.87.054901",
			year           = "2013",
			eprint         = "1212.1008",
			archivePrefix  = "arXiv",
			primaryClass   = "nucl-th",
			xSLACcitation   = "%%CITATION = ARXIV:1212.1008;%%",
}

@article{Qiu:2012uy,
			author         = "Qiu, Zhi and Heinz, Ulrich",
			title          = "{Hydrodynamic event-plane correlations in Pb+Pb
												collisions at $\sqrt{s}=2.76$ATeV}",
			journal        = "Phys.~Lett.~B",
			volume         = "717",
			pages          = "261",
			doi            = "10.1016/j.physletb.2012.09.030",
			year           = "2012",
			eprint         = "1208.1200",
			archivePrefix  = "arXiv",
			primaryClass   = "nucl-th",
			xSLACcitation   = "%%CITATION = ARXIV:1208.1200;%%",
}

@article{Teaney:2013dta,
			author         = "Teaney, D. and Yan, L.",
			title          = "{Event-plane correlations and hydrodynamic simulations of
												heavy ion collisions}",
			journal        = "Phys.~Rev.~C",
			volume         = "90",
			pages          = "024902",
			doi            = "10.1103/PhysRevC.90.024902",
			year           = "2014",
			eprint         = "1312.3689",
			archivePrefix  = "arXiv",
			primaryClass   = "nucl-th",
			xSLACcitation   = "%%CITATION = ARXIV:1312.3689;%%",
}

@article{Bozek:2010vz,
			author         = "Bozek, Piotr and Broniowski, Wojciech and Moreira, Joao",
			title          = "{Torqued fireballs in relativistic heavy-ion collisions}",
			journal        = "Phys. Rev. C",
			volume         = "83",
			year           = "2011",
			pages          = "034911",
			doi            = "10.1103/PhysRevC.83.034911",
			eprint         = "1011.3354",
			archivePrefix  = "arXiv",
			primaryClass   = "nucl-th",
			SLACcitation   = "%%CITATION = ARXIV:1011.3354;%%"
}

@article{Jia:2014jca,
			author         = "Jia, Jiangyong",
			title          = "{Event-shape fluctuations and flow correlations in
												ultra-relativistic heavy-ion collisions}",
			journal        = "J. Phys. G",
			volume         = "41",
			year           = "2014",
			number         = "12",
			pages          = "124003",
			doi            = "10.1088/0954-3899/41/12/124003",
			eprint         = "1407.6057",
			archivePrefix  = "arXiv",
			primaryClass   = "nucl-ex",
			SLACcitation   = "%%CITATION = ARXIV:1407.6057;%%"
}

@article{Jia:2014ysa,
			author         = "Jia, Jiangyong and Huo, Peng",
			title          = "{Forward-backward eccentricity and participant-plane
												angle fluctuations and their influences on longitudinal
												dynamics of collective flow}",
			journal        = "Phys.~Rev.~C",
			volume         = "90",
			pages          = "034915",
			doi            = "10.1103/PhysRevC.90.034915",
			year           = "2014",
			eprint         = "1403.6077",
			archivePrefix  = "arXiv",
			primaryClass   = "nucl-th",
			xSLACcitation   = "%%CITATION = ARXIV:1403.6077;%%",
}

@article{Jia:2017hbm,
			author         = "Jia, Jiangyong and Zhou, Mingliang and Trzupek, Adam",
			title          = "{Revealing long-range multiparticle collectivity in small
												collision systems via subevent cumulants}",
			journal        = "Phys. Rev. C",
			volume         = "96",
			year           = "2017",
			number         = "3",
			pages          = "034906",
			doi            = "10.1103/PhysRevC.96.034906",
			eprint         = "1701.03830",
			archivePrefix  = "arXiv",
			primaryClass   = "nucl-th",
			SLACcitation   = "%%CITATION = ARXIV:1701.03830;%%"
}

@article{Jia:2017kdq,
			author         = "Jia, Jiangyong and Huo, Peng and Ma, Guoliang and Nie,
												Maowu",
			title          = "{Observables for longitudinal flow correlations in
												heavy-ion collisions}",
			journal        = "J. Phys. G",
			volume         = "44",
			year           = "2017",
			pages          = "075106",
			doi            = "10.1088/1361-6471/aa74c3",
			eprint         = "1701.02183",
			archivePrefix  = "arXiv",
			primaryClass   = "nucl-th",
			SLACcitation   = "%%CITATION = ARXIV:1701.02183;%%"
}

@article{Aad:2008zzm,
			author         = "{ATLAS Collaboration}",
			title          = "{The ATLAS Experiment at the CERN Large Hadron Collider}",
			xcollaboration  = "ATLAS",
			journal        = "JINST",
			volume         = "3",
			pages          = "S08003",
			doi            = "10.1088/1748-0221/3/08/S08003",
			year           = "2008",
			SLACcitation   = "%%CITATION = JINST,3,S08003;%%",
}

@article{Aaboud:2016leb,
			author         = "{ATLAS Collaboration}",
			title          = "{Performance of the ATLAS trigger system in 2015}",
			collaboration  = "ATLAS",
			journal        = "Eur. Phys. J. C",
			volume         = "77",
			year           = "2017",
			number         = "5",
			pages          = "317",
			doi            = "10.1140/epjc/s10052-017-4852-3",
			eprint         = "1611.09661",
			archivePrefix  = "arXiv",
			primaryClass   = "hep-ex",
			reportNumber   = "CERN-EP-2016-241",
			SLACcitation   = "%%CITATION = ARXIV:1611.09661;%%"
}

@article{Aad:ID,
			author         = "{ATLAS Collaboration}",
			title          = "{The ATLAS Inner Detector commissioning and calibration}",
			journal        = "Eur.~Phys.~J.~C",
			volume         = "70",
			pages          = "787",
			doi            = "10.1140/epjc/s10052-010-1366-7",
			year           = "2010",
			eprint         = "1004.5293",
			archivePrefix  = "arXiv",
			primaryClass   = "physics.ins-det",
			xreportNumber   = "CERN-PH-EP-2010-043",
}

@article{Aad:2011yr,
			author         = "{ATLAS Collaboration}",
			title          = "{Measurement of the centrality dependence of the charged
												particle pseudorapidity distribution in lead-lead
												collisions at $\sqrt{s_{_{NN}}}=2.76$ TeV with the ATLAS
												detector}",
			xcollaboration  = "ATLAS",
			journal        = "Phys.~Lett.~B",
			volume         = "710",
			pages          = "363",
			doi            = "10.1016/j.physletb.2012.02.045",
			year           = "2012",
			eprint         = "1108.6027",
			archivePrefix  = "arXiv",
			primaryClass   = "hep-ex",
			xreportNumber   = "CERN-PH-EP-2011-118",
			xSLACcitation   = "%%CITATION = ARXIV:1108.6027;%%",
}

@article{Miller:2007ri,
			author         = "Miller, Michael L. and Reygers, Klaus and Sanders,
												Stephen J. and Steinberg, Peter",
			title          = "{Glauber Modeling in High-Energy Nuclear Collisions}",
			journal        = "Ann.~Rev.~Nucl.~Part.~Sci.",
			volume         = "57",
			pages          = "205",
			doi            = "10.1146/annurev.nucl.57.090506.123020",
			year           = "2007",
			eprint         = "nucl-ex/0701025",
			archivePrefix  = "arXiv",
			xSLACcitation   = "%%CITATION = NUCL-EX/0701025;%%",
}

@article{Gyulassy:1994ew,
			author         = "Gyulassy, Miklos and Wang, Xin-Nian",
			title          = "{HIJING 1.0: A Monte Carlo program for parton and
												particle production in high-energy hadronic and nuclear
												collisions}",
			journal        = "Comput.~Phys.~Commun.",
			volume         = "83",
			pages          = "307",
			doi            = "10.1016/0010-4655(94)90057-4",
			year           = "1994",
			eprint         = "nucl-th/9502021",
			archivePrefix  = "arXiv",
			xreportNumber   = "LBL-34246",
			xSLACcitation   = "%%CITATION = NUCL-TH/9502021;%%",
}

@article{Masera:2009zz,
			author         = "Masera, M. and Ortona, G. and Poghosyan, M.G. and Prino,
												F.",
			title          = "{Anisotropic transverse flow introduction in Monte Carlo
												generators for heavy ion collisions}",
			journal        = "Phys.~Rev.~C",
			volume         = "79",
			pages          = "064909",
			doi            = "10.1103/PhysRevC.79.064909",
			year           = "2009",
			SLACcitation   = "%%CITATION = PHRVA,C79,064909;%%",
}

@article{Agostinelli:2002hh,
			author         = "Agostinelli, S. and others",
			title          = "{GEANT4 -- a simulation toolkit}",
			collaboration  = "GEANT4",
			journal        = "Nucl.~Instrum.~Meth.~A",
			volume         = "506",
			pages          = "250",
			doi            = "10.1016/S0168-9002(03)01368-8",
			year           = "2003",
			reportNumber   = "SLAC-PUB-9350, FERMILAB-PUB-03-339",
			SLACcitation   = "%%CITATION = NUIMA,A506,250;%%",
}
 
\clearpage
 
\begin{flushleft}
\hypersetup{urlcolor=black}
{\Large The ATLAS Collaboration}

\bigskip

\AtlasOrcid[0000-0002-6665-4934]{G.~Aad}$^\textrm{\scriptsize 102}$,    
\AtlasOrcid[0000-0002-5888-2734]{B.~Abbott}$^\textrm{\scriptsize 128}$,    
\AtlasOrcid{D.C.~Abbott}$^\textrm{\scriptsize 103}$,    
\AtlasOrcid[0000-0002-2788-3822]{A.~Abed~Abud}$^\textrm{\scriptsize 36}$,    
\AtlasOrcid[0000-0002-1002-1652]{K.~Abeling}$^\textrm{\scriptsize 53}$,    
\AtlasOrcid[0000-0002-2987-4006]{D.K.~Abhayasinghe}$^\textrm{\scriptsize 94}$,    
\AtlasOrcid[0000-0002-8496-9294]{S.H.~Abidi}$^\textrm{\scriptsize 167}$,    
\AtlasOrcid[0000-0002-8279-9324]{O.S.~AbouZeid}$^\textrm{\scriptsize 40}$,    
\AtlasOrcid{N.L.~Abraham}$^\textrm{\scriptsize 156}$,    
\AtlasOrcid[0000-0001-5329-6640]{H.~Abramowicz}$^\textrm{\scriptsize 161}$
% ...
\bigskip
\\

% ...
$^{36}$CERN, Geneva; Switzerland.\\
% ...
$^{40}$Niels Bohr Institute, University of Copenhagen, Copenhagen; Denmark.\\
% ...
$^{53}$II. Physikalisches Institut, Georg-August-Universit\"{a}t G\"ottingen, G\"ottingen; Germany.\\
% ...
$^{94}$Department of Physics, Royal Holloway University of London, Egham; United Kingdom.\\
% ...
$^{102}$CPPM, Aix-Marseille Universit\'e, CNRS/IN2P3, Marseille; France.\\
$^{103}$Department of Physics, University of Massachusetts, Amherst MA; United States of America.\\
% ...
$^{128}$Homer L. Dodge Department of Physics and Astronomy, University of Oklahoma, Norman OK; United States of America.\\
% ...
$^{156}$Department of Physics and Astronomy, University of Sussex, Brighton; United Kingdom.\\
% ...
$^{161}$Raymond and Beverly Sackler School of Physics and Astronomy, Tel Aviv University, Tel Aviv; Israel.\\
% ...
$^{167}$Department of Physics, University of Toronto, Toronto ON; Canada.\\
% ...

\end{flushleft}


\end{document}